*Why the US Needs a Deep Domestic Research Facility: <u>Owning</u> rather than Renting the Education Benefits, Technology Advances, and Scientific Leadership of Underground Physics*

Kevin Lesko

LBNL

**Introduction and Background**

In December 2010 the Committee on Plans and Progress of the National Science Board (NSB), which oversees the NSF, voted to cease planning activities on the Deep Underground Science and Engineering Laboratory (DUSEL). Discussions between the DOE and the NSF had advanced to the point of creating an advanced draft Memorandum of Understanding between the agencies outlining the stewardship of the major science efforts and outlining the approximate financial requirements between the agencies. The largest share of the proposed DUSEL construction funds devoted to supporting the experiments was allocated to support of Dark Matter, Neutrinoless Double Beta Decay, Long Baseline Neutrinos, and Nuclear Astrophysics. The cancellation of planning for DUSEL disrupted the planning for these experiments.

The DOE Office of Science acted expeditiously to preserve the Homestake site while it considered options for the DOE-supported experiments including LUX (Dark Matter), Majorana Demonstrator (Neutrinoless Double-Beta Decay), and the major HEP neutrino experiment, LBNE. The National Academy Report on DUSEL science [1], completed shortly after the NSB action, highlighted the critical nature of underground science to our understanding of the universe and opportunities for major discoveries. This report and the Office of Science sponsored report investigating options for the DOE [2], noted the positive impact that hosting a deep underground facility in the US would have on US leadership of science and training of future students.

The NRC report concluded regarding the importance of underground physics experiments:

*"Three underground experiments to address fundamental questions regarding the nature of dark matter and neutrinos would be of paramount and comparable scientific importance:*

    *-The direct detection dark matter experiment,*

    *-The long baseline neutrino oscillation experiment, and*

    *-The neutrinoless double-beta decay experiment.*

*Each of the three experiments addresses at least one crucial unanswered question upon whose answer the tenets of our understanding of the universe depend."*

The NRC report concluded regarding leadership opportunities and the benefits of a domestic facility:

*"The co-location of the three main underground physics experiments at a single site would be a means of efficiently sharing infrastructure and personnel and of fostering synergy among the scientific communities. The infrastructure at the site would also facilitate future underground research, either as extensions of the initial research program or as new research initiatives.*

*These added benefits, along with the increase in visibility for U.S. leadership in the expanding field of underground science, would be important considerations when siting the three physics experiments."*

The Office of Science Report concluded regarding the scientific and leadership opportunities:

*"There are compelling scientific motivations for all three experiments [dark matter, neutrinoless double beta decay and long baseline neutrinos] and important opportunities for the US to take and maintain leadership positions by creating the facility in the US.*

*Locating the facility in the U.S. would help to promote U.S. leadership in these fields for the foreseeable future*

*The time needed to carry out the three experiments will extend over two decades or more from now, including about one decade before data taking begin. In each case it is quite likely that there will be upgrades and follow-on experiments that will further extend the time scale of these physics programs*

*Given the extent of investment needed to carry out these experiments, the long timescales and the likelihood of follow-on experiments in each of these areas of research, the committee recognizes that there are major advantages to developing a common underground site for these experiments. Advantages include:*
- *Opportunities to share expensive infrastructure and to coordinate design efforts, construction, management and operations.*
- *Significant benefits in training the next and subsequent generations of scientists by having a common facility serve as an intellectual center in these fields of research. "*

These conclusions remain in force, unaltered by the NSF's decision on DUSEL.

**Creation of the US's Deep Underground Facility, SURF**

In Fiscal Year 2012 the DOE's Offices of HEP and NP funded the operations of the Sanford Underground Research Facility (SURF), establishing an operations office at LBNL and supporting the continued activities at SURF through a contract between LBNL and the South Dakota Science and Technology Authority (SDSTA). The supported activities include:

- the continued preservation of basic infrastructure including continued removing the inflow of water from the facility,

- maintaining safe access to the underground for preserving the infrastructure and also to support the science program sponsored by SURF in the Davis Campus (4850 feet-underground), and

- facility operations in support of the early science program, consisting of the LUX and Majorana Demonstrator experiments.

2012 saw the completion of the Davis Campus. SDSTA financed and led this major undertaking with state funds and the T. Denny Sanford donation. In total the SDSTA invested over $120M in the facility, collaborating with the DUSEL project to develop the Preliminary Design Report [3] as well as creating the Sanford Laboratory. The Sanford donation is exceptional – it is an investment of $70M by a philanthropist in direct support of experiments, with the creation of the Sanford Lab and infrastructure necessary for multiple experiments. The laboratory space will support at least a decade of physics experiments without significant upgrades. The SDSTA is funding the rehabilitation of the Ross Shaft to ensure it is ready for major new construction that would be needed for an underground long baseline neutrino experiment, a generation-3 dark matter experiment, or a one-tonne neutrinoless double-beta decay experiment. In March 2013, the state allocated another $2M towards the Ross effort and nearly $2M for establishment of a physics PhD program in the state funding approximately eight new faculty physics.

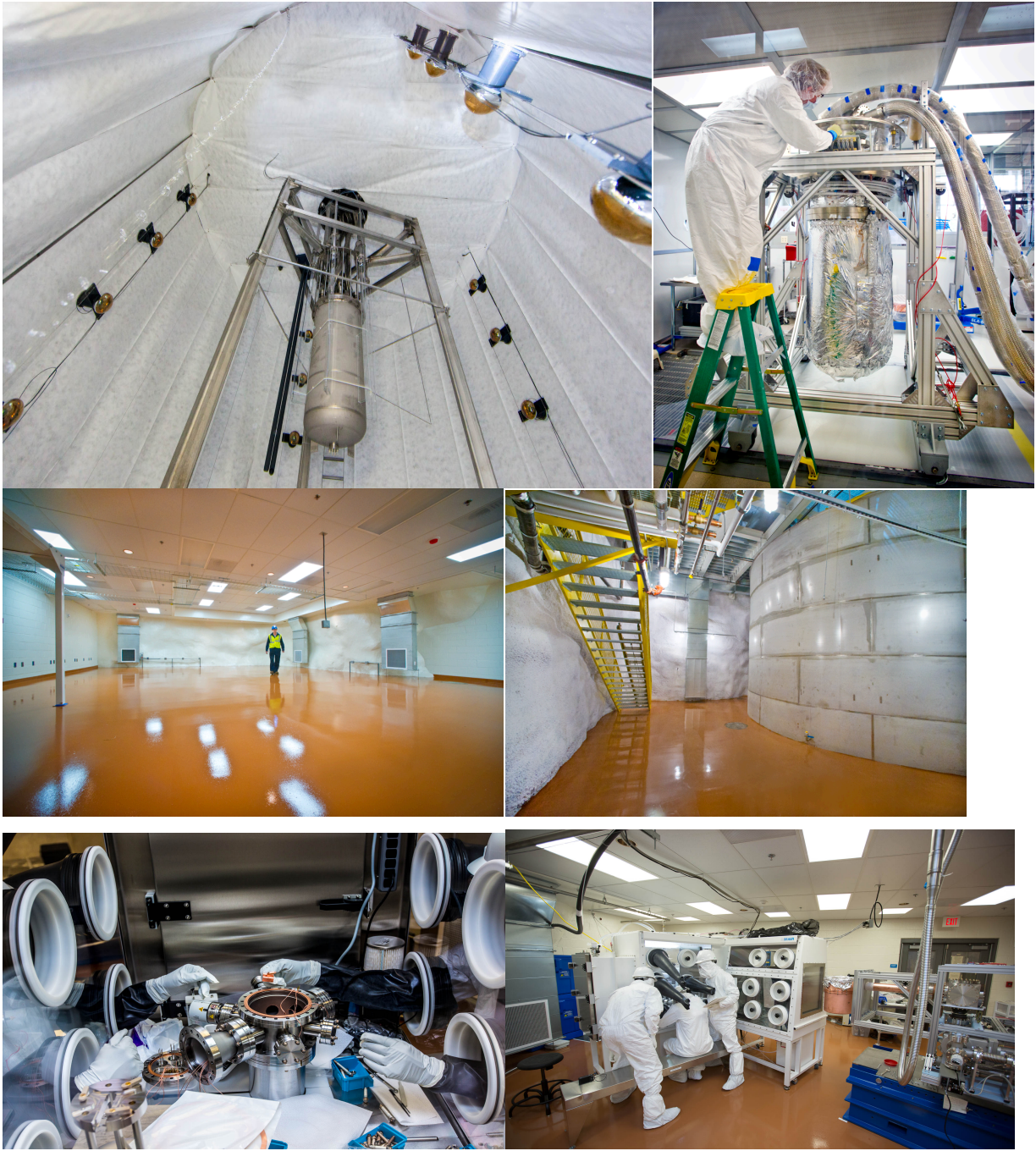

*Figure 1 The Sanford Underground Research Facility.  Top Row: Left - LUX detector installed, right – LUX detector being prepared.  Middle Row: Left - Majorana Demonstrator Laboratory, Right – LUX Water Shielding Tank.  Bottom Row: Assembling Majorana Detectors.*

The Davis campus provides the laboratory infrastructure needed to host these state-of-the-art experiments in the deepest underground laboratory in the US, and one of a small number of laboratories world-wide providing ~ 4300 mwe of shielding. There exists space for additional research efforts at the 4850 Level. The NSF's DIANA project, designed to create an underground nuclear astrophysics facility is investigating the 4850 Level to host their accelerator facility, in a well-isolated location.

In FY13 DOE HEP assumed the full responsibility of the facility operations while LUX and Majorana completed their construction activities and commenced detector commissioning. First results from both of these experiments are expected this year. LBNE has received CD-1 authorization and is advancing their designs for both the surface and underground options of the far detector.

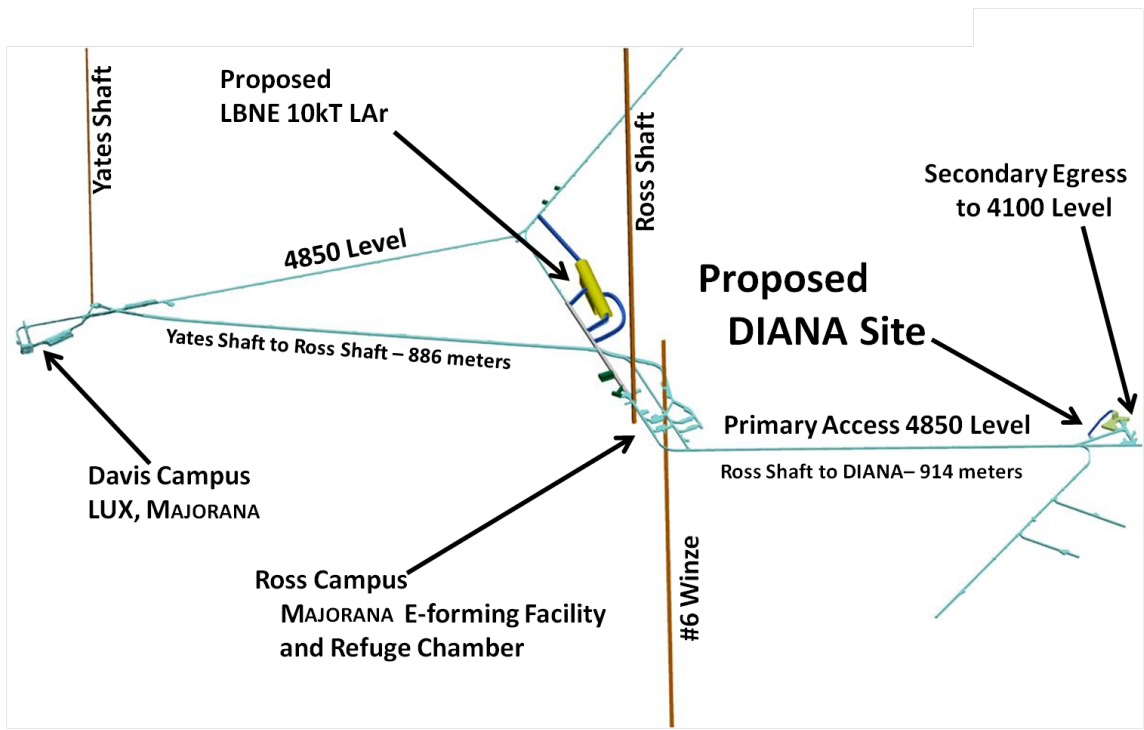

Figure 2 *LBNE underground options at SURF along with the proposed DIANA project and the Davis Campus. Proposed Generation-3 Laboratory Modules would be located between the Ross and Yates Shaft (not shown).*

With DOE's Office of Science leadership and the steadfast support of South Dakota, the US has established an initial operational underground research facility. SURF is on track to host all three of the most critical experimental experiments: Dark Matter, Neutrinoless Double Beta Decay, and Long Baseline Neutrinos. The successful development approach used to complete the current SURF infrastructure provides a flexible model for future and incremental growth for larger experiments. This approach addresses the shortcomings the large, all-at-once, plan envisioned by the NSF for DUSEL and is responsive the budget realities facing the Office of Science Programs. SURF can support phased programs and introduction of additional experiments and efforts.

**Reasons to Maintain a Domestic Deep Underground Research Facility and Support US Experiments at SURF**

**Scientific Leadership** – As concluded by the NRC and the Office of Science reports, operation of a deep domestic facility significantly enhances the international visibility of the US science program and establishes an avenue for continued US leadership in the science. The modest investments in the facility enable and encourage strong investments in our science. The US science collaborations will set the pace for the science. Foreign collaborators will bring essential contributions to the experiments, while the US teams will be leading the development of the science programs.

**Advanced Technology** – The technology required for next-generation experiments requires ready access to underground facilities to be developed. Space well-shielded from cosmic rays and environmental backgrounds is required in the development of detectors. A deep facility with adequate access and infrastructure is key for continuing the programs in dark matter searches and neutrino studies. There is a worldwide shortage of appropriate underground space. SURF can support additional modest R&D efforts and has a demonstrated ability to expand to meet additional needs.

**Education and Public Outreach** – A domestic facility very significantly enhances the science community's effort to train the next generation of scientists. The overheads to station students and post-docs at a domestic facility are very substantially less than those required to send them abroad. Importantly, a domestic facility significantly enhances the re-assimilation of the students and post-docs into the US science markets. The experimental focus will be in the US, rather than in Europe, Asia or Canada, and the next generation of scientists will seek employment in the US where they can lead future research efforts, rather than seeking employment at foreign universities and laboratories as has been seen with the HEP programs centered off-shore.

**Control of Our Scientific Future** – We control our scientific future by establishing the facility and investing in the critical infrastructure. The US agencies will establish the support and ensure that it is maintained at a level adequate for our scientific goals and appropriate for our schedules. We are not dependent on foreign entities or agencies to create and control the environment that is required for the experiments. If the foreign laboratory is too shallow for next generation experiments, we won't have to research and attempt to develop expensive cosmic-ray tagging devices to compensate for this shortcoming. A dedicated facility like SURF is not compromised by the interference of foreign mining influences or transportation requirements that restrict access and compromise science at other facilities. Most importantly, a domestic facility provides the US Congress and our funding agencies a facility and program they can defend when seeking funds for the science. US investment in a domestic underground research facility is an investment that the local congressional delegations will fight for, and that all of Congress can defend. In contrast, US PIs hosting an event in a foreign embassy raises questions in Congress about their support for the fundamental science. Congress is well within their rights to ask: "Why isn't this done in the US where we will benefit directly from the investment?"

**Conclusion – Rent or Own?**

These four reasons can be summarized by a simple question: Do we want to rent these critical elements or would we rather own and lead our science long term? There many examples of how the US science community benefits from investments in research infrastructure. These investments enable scientific leadership on a sustainable path. The investment fosters research and advance technology development. A domestic facility strongly advances the training of the next generation of scientists. While using of a foreign facility may benefit a few collaborations over the next several years, exporting our best students, our technology, and outsourcing scientific leadership does not establish a long-term, sustainable program of transformational research and scientific advances. Far from taking away from the scientific efforts, the investments in a domestic underground facility will amplify the support for US science and the experiments. The development of SURF has proven that a domestic facility can be advanced in an incremental, phased way that is sensitive to the budget realities facing US-funded science efforts. We have established a deep domestic research facility. Will we capitalize on achieving this milestone and focus domestic underground research efforts at SURF including important foreign contributions to the US efforts or will we offshore our efforts and students to foreign facilities and cede US scientific leadership?

---